\documentstyle[12pt,psfig,epsfig]{article}
\newcommand{\beq}{\begin{equation}}
\newcommand{\eeq}{\end{equation}}
\newcommand{\beqarray}{\begin{eqnarray}}
\newcommand{\eeqarray}{\end{eqnarray}}
\def\lsim{\raise0.3ex\hbox{$\;<$\kern-0.75em\raise-1.1ex\hbox{$\sim\;$}}}
\def\gsim{\raise0.3ex\hbox{$\;>$\kern-0.75em\raise-1.1ex\hbox{$\sim\;$}}}
\def\para{\vspace{0.3cm}\noindent}

\def\km{\,{\rm km}}

\def\mpc{\,{\rm Mpc}}

\def\sec{\,{\rm sec}}

\begin{document}
\begin{center}
{\large \bf Neutrino Induced Upward Going Muons from a Gamma Ray  \\
  Burst in a Neutrino Telescope of $Km^{2}$ Area}

\medskip

{Nayantara Gupta \footnote{tpng@mahendra.iacs.res.in}}\\  
{\it Department of Theoretical Physics,\\
 Indian Association for the Cultivation of Science,\\
Jadavpur, Kolkata 700 032, INDIA.}  

\end{center}

\begin{abstract}
The number of neutrino induced upward going muons from a single gamma ray burst
(GRB) expected to be detected by the proposed kilometer scale IceCube detector at the South Pole location has been calculated.
  The effects of the Lorentz factor, total energy of the GRB emitted in neutrinos and its distance from the observer (redshift)  on the number of neutrino events from that GRB have been examined. The present investigation reveals that there is possibility of
exploring physical processes of the early Universe with the proposed kilometer
scale IceCube neutrino telescope. 
\end{abstract} 

PACS number(s): 96.40.Tv, 98.70.Rz, 98.70.Sa 
\newpage
\section{Introduction}
We consider the relativistically expanding fireball model of
 gamma ray bursts (GRB) \cite{Eli} and calculate the number of secondary muons from high energy neutrinos of a GRB expected to be detected in a kilometer
scale neutrino telescope. 
In a relativistic expanding fireball protons may be accelerated by Fermi mechanism to energies as high as  $10^{20} eV$.
The accelerated protons interact with photons of energies of the order of MeV to produce pions.
From protons both $\pi^{+}$'s and $\pi^{0}$'s are produced sharing the
proton energies roughly in equal parts. The $\pi^{+}$'s would lead to generation of high energy neutrinos 
and from $\pi^{0}$'s high energy photons would be produced.

The decay process for neutrino production from $\pi^{+}$'s is as follows
$\pi^{+}\rightarrow\mu^{+}+\nu_{\mu}\rightarrow e^{+}+\nu_{e}+\bar\nu_{\mu}+\nu_{\mu}$.
The Lorentz factor $\Gamma$ of a GRB has a crucial role in the neutrino production
mechanism of the GRB. The efficiency for producing pions in $p-\gamma$ collisions in the fireball varies as $\Gamma^{-4}$ and the break energy for neutrino 
energy spectra varies as $\Gamma^{2}$. Although $\Gamma$ grows linearly with
the radius in a relativistic fireball model, $\Gamma$ saturates at a value
of the order of 100 \cite{Eli}.  

\para
The fraction of energy lost by protons to pions $f_{\pi}$ can be expressed as a
function of the pair production optical depth $\tau_{{\gamma}{\gamma}}$.
If the optical depth is large there would be more pion productions. Also if the optical depth is large then the photons produced from $\pi^{0}$ decays can not come out as a burst. In that case the GRB would be opaque for photons. The GRBs which are not bright in photons are bright in neutrinos.

\para
The effects of varying  the Lorentz factor $\Gamma$, the total
energy emitted or released in neutrinos $E_{GRB}$ and redshift of the GRB $z$ 
on the expected number of neutrino induced muons from a GRB have been
studied in this work. Each gamma ray burst is an individual phenomenon in
the Universe distinguished by its redshift, total energy emitted in neutrinos or photons, duration of the
burst and Lorentz factor at a particular instant of time .
To obtain informations on the physical process responsible for a gamma ray
burst it is always meaningful to carry out studies on the detectability
 of individual GRBs.

\para

In Ref. \cite{Guetta} the correlations of the parameters minimum Lorentz factor $\Gamma$, wind variability time $t_{v}$, observed photon spectral break energy 
 $E_{\gamma, MeV}^{b}$ for wind luminosity $L_{w}=10^{53}$ erg are given    
in FIG.1.for equipartition parameters $\epsilon_{B}=0.01$ and $\epsilon_{B}=0.1$.
We have taken points from the contour plots given in FIG.1. of Ref. \cite{Guetta} and used those values in our calculations.
In the present work we use the wind luminosity $L_{w}$ but by $E_{GRB}$ we mean the total energy released in neutrino emission.
The observed photon spectral break energy $E_{{\gamma},MeV}^{b}$ is the energy where luminosity per logarithmic photon energy interval peaks.  

The fraction of fireball energy that goes into neutrino production is weakly dependent on the wind model parameters. There are two reasons for that. 
The first reason is for low values of Lorentz factor $\Gamma$ and wind variability time $t_{v}$ only a small fraction of the pion's energy is converted to neutrinos at high proton energy due to pion and muon synchrotron losses. The second
reason is the observational constraints imposed by $\gamma$-ray observations imply that the wind model parameters $\Gamma$, wind luminosity $L_{w}$ and wind 
variability time $t_{v}$ are correlated. We keep in mind the luminosity mentioned in \cite{Guetta} is the wind luminosity and in this work also we use the
same notation.
To avoid confusion we mention that the fraction of proton energy that goes into
neutrino production does not appear explicitly in the expression of total number of neutrinos emitted from a GRB in our procedure of calculation. The reason behind this is we have used $E_{GRB}$ which is the total energy released in neutrinos to calculate the number of neutrinos emitted from a GRB.

\para
From one proton two muon neutrinos are expected to be generated if both the pion
 and the muon decay. The neutrinos carry about $5\%$ of the proton energy, it
is given in \cite{Guetta}. So the total energy $E_{GRB}$ released in 
neutrinos of a GRB is about $10\%$ of the original total fireball proton energy. We have used the total energy emitted in neutrinos $E_{GRB}$ to calculate 
the neutrino spectrum from a GRB. So the energy loss by protons in pion productions as well as the pion and muon energy losses which are the intermediate processes for neutrino productions from protons have been already taken into account in our procedure of calculations. 
The small value of $E_{GRB}$ compared to the original total fireball energy accounts for the intermediate energy loss processes in production of neutrinos from
 protons in a GRB. If we use $E_{GRB}=10^{53}$erg in our calculation that means the total fireball proton energy was $10^{54}$erg assuming $10\%$ of the fireball energy has been released in neutrino emission from the GRB. The wind durations are of the order of 10 seconds.

\para
The neutrino spectrum from a GRB on Earth depends on the 
total energy emitted in neutrinos $E_{GRB}$, distance of the GRB from the observer $z$ and the neutrino break energy  $E_{\nu}^{b}$ which is again  dependent on the Lorentz factor $\Gamma$, observed photon spectral break energy $E_{{\gamma},MeV}^{b}$, the wind variability time $t_{v}$ as well as wind luminosity $L_{w}$ and equipartition parameter $\epsilon_{B}$.

\para 
Earlier average neutrino event rates in a neutrino telescope of $km^{2}$ area per year from gamma ray bursts have been obtained 
considering burst-to-burst fluctuations in distance and energy \cite{hal,alv}.
They have used energy and redshift distribution functions of the GRBs  modeled according to observations on GRBs. The neutrino event rates per year from GRBs have been obtained including the fluctuations in the Lorentz factor $\Gamma$.
The purpose of the present work is not to calculate 
the neutrino event rate expected per year in a muon detector of $km^{2}$ area.
We investigate for what range of values of their physical parameters ($E_{GRB}$, $z$, $\Gamma$)  individual GRBs will be detectable by a kilometer scale neutrino telescope of a given threshold energy.  

\para
In GRBs $\nu_{\mu}+\bar\nu_{\mu}$'s are almost $10^{5}$ times more abundant
than $\nu_{\tau}+\bar\nu_{\tau}$. The primary neutrino flux from GRBs contain
very small number of $\nu_{\tau}+\bar\nu_{\tau}$.  
Due to neutrino oscillations some of the $\nu_{\mu}$ 's convert to $\nu_{\tau}$'s during their propagation from the burst location to the detector on Earth .
According to SuperKamiokande measurements \cite{sup} the flux ratio of
muon neutrinos and tau neutrinos is  $F_{\nu_{\tau}/\nu_{\mu}}=0.5$.
 The 
$\nu_{\tau}+\bar\nu_{\tau}$ events would produce different experimental signatures from the $\nu_{\mu}+\bar\nu_{\mu}$ events \cite{alv}.
Cosmic tau neutrinos while coming close to the surface of the detector may undergo a charged current deep inelastic scattering with nuclei inside or near the
detector and produce a tau lepton in addition to a hadronic shower. This 
tau lepton traverses a distance, on average proportional to its energy, before
it decays back into a tau neutrino and a second shower most often induced
by decaying hadrons. The second shower is expected to carry about twice as
much energy as the first and such double shower signals are commonly 
referred to as double bangs. 
The tau leptons produced in this way are not expected to have any other
relevant interactions as they are losing energy very fast and subsequently they decay to muons. 
Other than double bang events there are $\nu_{\tau}+\bar\nu_{\tau}$ events
in which muons would be detected from $\tau$ decays $(\nu_{\tau}\rightarrow
\tau\rightarrow\mu$). The double bang events are much less compared to the events  in which tau leptons decay to muons ($\nu_{\tau}\rightarrow\tau\rightarrow\mu$). The $\nu_{\mu}+\bar\nu_{\mu}$ events are much more in number compared to the
$\nu_{\tau}+\bar\nu_{\tau}$ events. 
 The  relative magnitudes of rates of double bang events from
 $\nu_{\tau}$'s, events in which $\tau$'s decay to $\mu$'s and the $\nu_{\mu}+\bar\nu_{\mu}$ events per year in a $km^{2}$ area neutrino telescope are given in
\cite{alv}. 
The majority of muon events in the $km^{2}$ area muon detector would be from interactions of
 $\nu_{\mu}+\bar\nu_{\mu}$ in rock (in case of upward going muons) or ice
 (in case of downward going muons) at the South Pole location. 
 
In this work we consider only the upward going muon events from $\nu_{\mu}+\bar\nu_{\mu}$ of a gamma ray burst because incase of upward going events the background noise is comparatively less.   

\para
Recent observations on GRB afterglow have confirmed the relativistic fireball model, to be more specific the internal-external shocks model \cite{piran}. If GRB neutrinos could be observed by neutrino telescopes it would be a great achievement for understanding the distribution in Lorentz factor $\Gamma$, the ultrahigh energy neutrino background, testing special relativity and the equivalence principle.

\para
In section 2 we discuss about the parametrization of $\nu_{\mu}+\bar\nu_{\mu}$
spectrum from a GRB and the procedure used here to obtain the number of neutrino induced muon signals. In section 3 we mention about the atmospheric neutrino
background which often makes the detection of neutrino signals from astrophysical sources difficult in muon detectors. In section 4 the results
of this work have been discussed.
\section{Parametrizations of the Neutrino Spectrum of a GRB and the 
Secondary Muons Produced in Rock}

\subsection{Neutrino Spectrum of a GRB}
A total amount of energy $E_{GRB}$ is released in neutrino emission from a
GRB at a distance of redshift $z$ from the observer on Earth. 

The Lorentz factor of the GRB is $\Gamma$.
The neutrino spectrum is expected to be correlated mainly with the observed
gamma ray spectrum of the GRB \cite{Guetta}. In analogy with observed $\gamma$-ray spectrum of a GRB we write down the neutrino spectrum of the GRB. 

\beq
\frac{dN_{\nu}}{dE_{\nu}} = A \times
min(1, E_{\nu}/E^{b}_{\nu})\frac{1}{E_{\nu}^{2}}
\eeq

Here $A$ is the normalisation constant and $E^{b}_{\nu}$ is the break energy of the neutrino spectrum which is a function of Lorentz factor $\Gamma$, photon
spectral break energy $E_{{\gamma}, MeV}^{b}$. Since $\Gamma$, $E_{{\gamma},MeV}^{b}$ and wind variability time $t_{v}$ are correlated so $E^{b}_{\nu}$ is also
dependent on $t_{v}$ for a given value of wind luminosity $L_{w}$ and equipartition parameter $\epsilon_{B}$.

\beq
E^{b}_{\nu}\approx 10^{6} \frac{{\Gamma}^{2}_{2.5}}{E^{b}_{{\gamma},MeV}} GeV
\eeq

In the above expression ${\Gamma}_{2.5}={\Gamma}/10^{2.5}$.
 $E_{{\nu}min}$ and $E_{{\nu}max}$ are the lower and upper
cutoff energy of the neutrino spectrum. We assume $E_{{\nu}min}<<E_{\nu}^{b}$. 

The normalisation constant A is determined from energy considerations
\beq
A=\frac{E_{GRB}}{1+\ln(\frac{E_{{\nu}max}}{E_{\nu}^{b}})}
\eeq   
 To make things clear we again mention that all the intermediate energy loss terms for neutrino productions from protons are accounted for within $E_{GRB}$ which is a small fraction of the initial fireball proton energy.
The observed neutrino energy on Earth $E_{{\nu}obs}$ and its energy at the source, $E_{\nu}$, are related as $E_{{\nu}obs}=E_{\nu}/(1+z)$.  
The observed cutoff energy of the source is
$E_{{\nu}maxobs}=E_{{\nu}max}/(1+z)$. We assume $E_{{\nu}max}=10^{11}
GeV$.
 Assuming isotropic
emission from the source, the total number
of neutrinos from a single GRB at redshift $z$ per unit energy at energy
$E_{{\nu}obs}$ that strike per unit area of the Earth 
 is 
\beq
\frac{dN_{{\nu}0}}{dE_{{\nu}obs}}=\frac{dN_{\nu}}{dE_{\nu}}\frac{1}{4\pi r^2(z)}
(1+z)
\eeq 

 $r(z)$ is the comoving radial coordinate distance of the source. 
For a spatially flat Universe (which we shall assume to be the case) with
$\Omega_{\Lambda} + \Omega_m = 1$, where 
$\Omega_{\Lambda}$ and $\Omega_m$ are, respectively, the
contribution of the cosmological constant and matter to the energy density
of the Universe in units of the critical energy density, $3H_0^2/8\pi G$, 
the radial coordinate distance $r(z)$ is given by  
\beq
r(z)=\int^{z}_{0}\left(\frac{c}{H_0}\right)\frac{dz^\prime}
{\sqrt{\Omega_{\Lambda}+ \Omega_m(1+z^\prime)^3}}\,.\label{r_z}
\eeq
Here $c$ is the speed of light, $H_0$ is the Hubble constant in the
present epoch. In our calculations, we shall use  
$\Omega_\Lambda=0.7$, $\Omega_m=0.3$, and $H_{0}=65\, \km\, \sec^{-1}\,
\mpc^{-1}$. A similar procedure for TeV photon flux calculation on Earth from GRBs has been used in \cite{pijush}.

\para
The orginal muon neutrinos from GRBs are undergoing oscillations during
propagation and tau neutrinos are produced in this way. 
The probablity of oscillation from $\nu_{\mu}$ to $\nu_{\tau}$ is given by
\beq
Prob(\nu_{\mu}\rightarrow\nu_{\tau})=sin^{2}2{\theta}sin^{2}(\frac{{\delta}m^{2}}
{4E_{\nu}d})
\eeq

Considering the values of neutrino mass difference and mixing as found from SuperKamiokande  data  $(sin^{2}2{\theta}\sim 1, {\delta}m^{2}\sim 10^{-3} eV^{2})$  \cite{sup} and the distance $d\sim 100 $ Mpc between AGN and our galaxy 
the above expression for probablity averages to about half for all relevant neutrino energies to be considered for detection \cite{athar}.  
The high-energy muon neutrino flux is suppressed by a costant factor of half independent of its energy due to neutrino oscillations. 
The expression for muon neutrino flux from GRBs we have obtained above has been divided  by a constant factor of two to include the effect of neutrino oscillations.
\subsection{Number of Secondary Muons Produced from Muon Neutrinos of a GRB} 
The muon events from GRB neutrinos have been calculated including muon energy
loss in rock and absorption of neutrinos during their passage through the
Earth.  
 The number of neutrino induced muons from a GRB of energy $E_{GRB}$, redshift
$z$ and Lorentz factor $\Gamma$ has been calculated above a muon threshold energy
 $E_{th}$ using the standard procedure from \cite{gaisser}.
The total number of secondary muons from the muon neutrinos of a GRB below the detector above a threshold energy $E_{th}$ is given by
\beq
S(>E_{th})=\int_{E_{th}}^{\infty}dE_{{\nu}obs}\frac{dN_{{\nu}0}}{dE_{{\nu}obs}}
P_{\nu}(E_{{\nu}obs},>E_{th})exp[-\sigma_{tot}(E_{{\nu}obs})N_{A}X(\theta)]
\eeq
In the above expression $\sigma_{tot}$ is the total cross section for neutrino
absorptions. It is the sum of charged-current and neutral-current cross sections
\cite{raj}. $X(\theta)$ is the matter traversed in $gm/cm^{2}$ by the neutrinos
 emitted from the GRB during their passage through the Earth at a zenith angle
 $\theta$. $N_{A}=6.022\times10^{23}cm^{-3}$.
The probablity that a muon produced in a charged-current neutrino nucleon
interaction arrives the underground detector with an energy above the muon
 threshold energy $E_{th}$ is 
$
P_{\nu}(E_{{\nu}obs}, >E_{th})
$.
It depends on the neutrino nucleon charged-current cross section of muon production and the energy loss rate of muons in rock. We have used the neutrino-nucleon interaction cross sections from \cite{raj}. The rate of muon energy loss in rock has been used from \cite{dar}.
 The expression for probablity of muon production from neutrinos is 
\beq
P_{\nu}(E_{{\nu}obs}, >E_{th})=\int_{E_{th}}^{E_{{\nu}obs}}dE_{\mu}\frac{{\xi} N_{A}}{(\epsilon+E_{\mu})}\int_{0}^{1-E_{\mu}/E_{{\nu}obs}}dy^{\prime}\int_{0}^{1}dx\frac{d\sigma}{dxdy^{\prime}}
\eeq
The charged-current differential cross sections can be expressed in power laws in neutrino energies.
\beq
\frac{d\sigma(E_{{\nu}obs})}{dE_{{\nu}obs}}=BE_{{\nu}obs}^{\delta}
\eeq
where $B$ is the amplitude and $\delta$ is the differential index of power law.
The expression of probablity of muon production from neutrinos can be simplified to
\beq
P_{\nu}(E_{{\nu}obs}, >E_{th})=\frac{{\xi} N_{A}B}{{\delta}+1}\int_{E_{th}}^{E_{{\nu}obs}}\frac{dE_{\mu}}{(\epsilon+E_{\mu})} E_{{\nu}obs}^{{\delta}+1}[1-(\frac{E_{\mu}}{E_{{\nu}obs}})^{\delta+1}]
\eeq

The expression for rate of energy loss by muons is
\beq
\frac{dE_{\mu}}{dx}=-\alpha-\beta E_{\mu}
\eeq
where for rock 
\beq
\alpha= [2.033+ 0.077 \ln E_{\mu}(GeV)]10^{-3}GeVgm^{-1}cm^{2}
\eeq
 and 
\beq
\beta=[2.229+0.2\ln E_{\mu}(GeV)]10^{-6}gm^{-1}cm^{2}
\eeq
 from \cite{dar}. In the expression of probablity
$\epsilon={\alpha}/{\beta}$ and $\xi=1/\beta$. 

The expression for probablity has been substituted in equation (8) and the number of neutrino induced muons above muon threshold energy $100  GeV$ has been calculated for a detector of $km^{2}$ area.

 \section{Atmospheric Neutrino Induced Muons in a Neutrino Telescope of $km^{2}$ Area}
 The atmospheric neutrino fluxes from \cite{agrawal} and the neutrino nucleon interaction cross sections from \cite{raj} have been used to calculate the neutrino induced muon fluxes from atmospheric neutrinos in a muon detector of $km^{2}$ area above a muon threshold energy $E_{th}=100 GeV$. The secondary muon fluxes produced in rock have been calculated for $\cos{\theta}=-1$ and $\cos{\theta}=-0.05$ where $\theta$ is the zenith angle .
 The procedure for calculation of neutrino induced muons as discussed
in section 2 is used in this case also including the absorption of neutrinos during their propagation through the Earth. The IceCube detector would be placed
at a depth of 2 km below the surface of Earth. We have used the effective  matter traversed by neutrinos inside the Earth when calculating the neutrino events for $\cos{\theta}=-0.05$.  

\section{Results and Discussions}
In FIG.1.of Ref.\cite{Guetta} contour plots of observed photon spectral break energy are given for $L_{w}=10^{53}$erg/sec and equipartition parameter $\epsilon_{B}=0.1$, $\epsilon_{B}=0.01$. 
From that figure for a value of minimum Lorentz factor $\Gamma$ we can find the allowed value of $E_{{\gamma}, MeV}^{b}$ and the corresponding value of wind variability time $t_{v}$ sec for $L_{w}=10^{53}$erg/sec and equipartition parameter $\epsilon_{B}=0.01$. For
one value of $\Gamma$ there can be many allowed values $E_{{\gamma}, MeV}^{b}$.
The values of minimum Lorentz factor $\Gamma$, wind variability time $t_{v}$ and observed photon spectral break energy $E_{{\gamma}, MeV}^{b}$ used in our work for wind luminosity $L_{w}=10^{53}$erg/sec and equipartition parameter $\epsilon_{B}=0.01$ have been listed in TABLE I.

\newpage
\begin{center}
TABLE I\\
The values of minimum Lorentz factor $\Gamma$, $t_{v}$ in seconds, $E_{{\gamma}, MeV}^{b}$ in MeV used in the present work for wind luminosity $L_{w}=10^{53}$ erg/sec and equipartition parameter $\epsilon_{B}=0.01$ from the contour plots
given in \cite{Guetta}.
\vskip 0.5cm
\begin{tabular}{|c|c|c|}
\hline
$\Gamma$ & $t_{v}$ sec & $E_{{\gamma}, MeV}^{b}$ \\
\hline
50.12&0.707&0.794\\
56.23&0.316&0.794\\
70.79&1.&3.162\\
79.43&0.316&3.162\\
89.12&0.1&3.162\\
100.&0.421&7.943\\
112.2&0.316&19.952\\
125.892&0.316&125.892\\
141.25&0.251&125.892\\
158.48&0.63&79.432\\
177.82&0.251&79.432\\
199.52&0.501&50.118\\
223.872&0.125&50.118\\
251.188&0.1&31.622\\
281.838&0.398&19.952\\
316.227&0.079&19.952\\
354.81&0.63&7.943\\
398.107&0.199&7.943\\
\hline
\end{tabular}
\end{center}
\vskip 1cm

\para

 The mean isotropic equivalent energy released by GRBs is about $10^{53}$erg from \cite{frail}. The total wind energy is related to the energy $E_{GRB}$ emitted in neutrinos by the GRB. If the energy released in neutrino emission $E_{GRB}$ is $10^{53}$ erg then the wind
energy is expected to be of the order of $10^{54}$erg since neutrinos carry about $5\%$ of proton's energy and from one proton two muon neutrinos are expected
to be produced.

\para
In Fig.1. of the present work the number of secondary muons from a GRB in $km^{2}$ area detector 
has been shown for different values of the Lorentz factors $\Gamma$ and $E_{{\gamma}, MeV}^{b}$. The energy emitted by the GRB in neutrinos is $10^{53}$erg and the burst is at zenith angle $\theta=180^{\circ}$. We have used the value of the threshold energy of IceCube detector
$E_{th}=100 GeV$. One can calculate the neutrino signals for higher threshold
energies of the detector. The  values of redshift $z$ of the GRB used in the calculation are 0.05,0.5,1. and 1.5.  
The Lorentz factor $\Gamma$ and the observed photon spectral break energy $E_{{\gamma}, MeV}^{b}$ together determine the value of the neutrino break energy $E_{\nu}^{b}$.

 If $\Gamma$ is increased the break energy of neutrino spectrum $E_{\nu}^{b}$ also
shifts to higher value, if the photon spectral break energy does not increase. The neutrinos produced from a GRB of high Lorentz factor and low $E_{{\gamma}, MeV}^{b}$ 
will have their energies higher than those produced from a GRB of lower Lorentz factor and higher $E_{{\gamma},MeV}^{b}$.
The atmospheric background noise decreases for higher threshold energy of the detector. So it will be easier for a detector of higher threshold energy to detect these GRB neutrinos emitted from GRBs of high Lorentz factors and low photon spectral break energies.

\para
In Fig.2. of this paper the effect of varying the redshift of the GRB on the production of 
neutrino events in the IceCube detector has been studied. As stated earlier
the burst is at zenith angle $\theta=180^{\circ}$, total energy emitted
 in neutrino production is $E_{GRB}=10^{53}$erg, wind luminosity of the GRB is $10^{53}$erg/sec and threshold of the detector
is $E_{th}=100 GeV$. If the distance of the GRB from the observer is large the
observer receives very few signals.
In Fig.3. of this paper the numbers of neutrino induced muons have been calculated for different redshifts when the GRB is at zenith angle $\theta=92.86^{\circ}$. The rest of the parameters are same as in Fig.2.

\para
In Fig.4. of the present paper the dependence of secondary muon events on the energy of the GRB $E_{GRB}$ has been shown.
If the total energy released by neutrino emission $E_{GRB}$ is increased obviously there will be more neutrino generation. The number of secondary muons from the GRB increases linearly with $E_{GRB}$.The wind luminosity of the GRB is assumed to be $L_{w}=10^{53}$erg/sec and redshift of the GRB is $z=1.5$. 

\para
Using the procedure discussed in section 3 we have calculated the atmospheric
neutrino generated background muon events per second in IceCube detector for
threshold energy $E_{th}=100 GeV$.
In case of $\cos{\theta}=-1$ we find 0.018 atmospheric  neutrino induced muon events from $\nu_{\mu}+\bar\nu_{\mu}$ per second per $km^{2}$ area and for $\cos{\theta}=-0.05$ the corresponding number is 0.04.

We denote the duration of time for neutrino emission from a GRB by $t_{\nu}$ at the source in observer's reference frame.
From Fig.2. and Fig.3. of this paper it is clear that a nearby burst of a few seconds duration ($t_{\nu}$) with energy $E_{GRB}$ of the order of $10^{53}$erg will be always detectable in IceCube for threshold energy $E_{th}=100GeV$.
\para

We have also investigated for a known redshift $z$ and Lorentz factor $\Gamma$ of the GRB
for what range of values of the total energy $E_{GRB}$ the IceCube detector with$E_{th}=100GeV$ will be able to detect the GRB at a zenith angle $\theta$.
By detectable we mean the number of secondary muons produced by the GRB is more than the background atmospheric neutrino induced muon events integrated over the observed duration of the burst.
If the duration of neutrino emission by the GRB is long the background noise is more compared to the case when the duration of neutrino emission is short. In calculating the observed duration (duration for neutrino emission ) of the burst on Earth the correction due to the redshift of the source has been taken into account. The observed duration of the GRB from Earth for neutrino emission is
$t_{\nu}\times (1+z)$ where $z$ is the redshift of that GRB.
We have calculated number of neutrino events from a GRB for $\cos{\theta}=-1$ and $\cos{\theta}=-0.05$ and duration of neutrino emission from the GRB ($t_{\nu}$ ) is assumed to be 1.second. The correction on the matter traversed by neutrinos due to 2km depth of the detector below the surface of the Earth has been taken into account.
In Fig.5. of this paper the regions above
the curves are the allowed regions of parameters ($\Gamma$ and $E_{GRB}$) for detectable GRBs at redshift $z=1.5$ in IceCube detector. One can extend this study for any redshift $z$, zenith angle of observation $\theta$ and duration of neutrino emission ($t_{\nu}$) from the GRB.

\para
Simultaneous observations of photons and neutrinos from a burst in a small time window will confirm its signature.
The GRBs are distributed uniformly as standard candles throughout the Universe
\cite{tsvi}. There is a large number of faint GRBs from which less than one event will be detected on average in the muon detector.

From redshift $z=1.5$ for $\Gamma=100$ and $E_{GRB}=10^{56}$erg we can expect
14.4 muon events for $E_{\gamma}^{b}=7.943MeV$ at zenith angle $\theta=180^{\circ}$ in IceCube neutrino telescope for threshold energy $E_{th}=100GeV$.
Even from a distance of redshift $z=2.1$ if total energy emitted in GRB neutrinos is $10^{56}$erg we can expect 8 muon events in IceCube for threshold energy $E_{th}=100 GeV$ at zenith angle $\theta=180^{\circ}$ when $\Gamma=100$ and $E_{\gamma}^{b}=7.943MeV$. 

So there is possibility of neutrino detection from bursts at high redshifts 
if they emit a large amount of energy in neutrinos $E_{GRB}$ in a short duration of time ($t_{\nu}$).  

\section{Conclusion}
We have investigated on the capability of the next generation kilometer scale
IceCube detector in detecting high energy neutrinos from individual GRBs.
It is always easier to detect nearby GRBs. Although the neutrino events steeply fall with increasing redshift of the GRB there is possibility of neutrino detection even from the far away bursts in IceCube neutrino telescope. If the GRB is of short duration $t_{\nu}$, low Lorentz factor $\Gamma$, emit large amount of energy $E_{GRB}$ in neutrino emission and if its photon  spectral break energy $E_{{\gamma}, MeV}^{b}$ is high the possibility of detecting neutrinos from that GRB increases. Not only understanding the physics of GRBs, testing special relativity and equivalence principle but we can also look forward to study physical processes occuring in the early Universe with neutrino telescope of $km^{2}$ area. 

\section{Acknowledgement} 
I wish to thank Pijushpani Bhattacharjee  for fruitful discussions, who inspired me to work on the detectability of individual gamma ray bursts. I am also
thankful to the referee for finding out inadequacies in the initial versions
of this paper.

\begin{figure}
\newpage
\psfig{figure=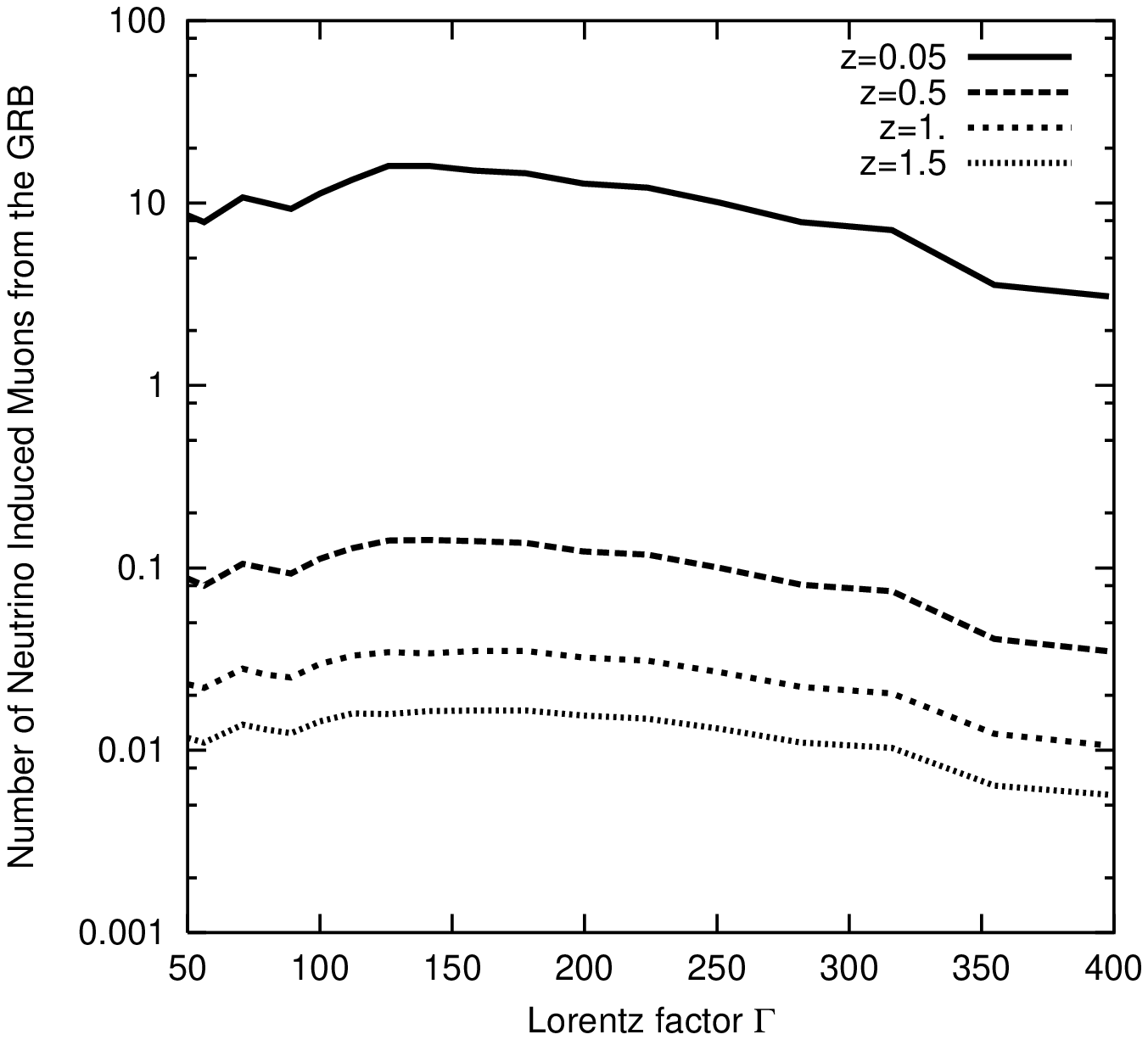, height=11cm} 
\caption{Number of secondary muons in IceCube detector from neutrinos produced from a GRB as a function of the Lorentz factor $\Gamma$ of the GRB have been plotted in this figure. The detector area is $1 km^{2}$ and the GRB is at a zenith angle of $\theta=180^{\circ}$. The wind luminosity of the GRB is assumed to be $10^{53}$erg/sec for equipartition parameter $\epsilon_{B}=0.01$. The GRB parameters $\Gamma$, observed variability time $t_{v}$ sec, phton spectral break energy $E_{{\gamma},MeV}^{b}$ used here have been given in TABLE I. The total energy released in neutrino emissions is assumed to be $10^{53}$erg.}
\end{figure}

\begin{figure}
\psfig{figure=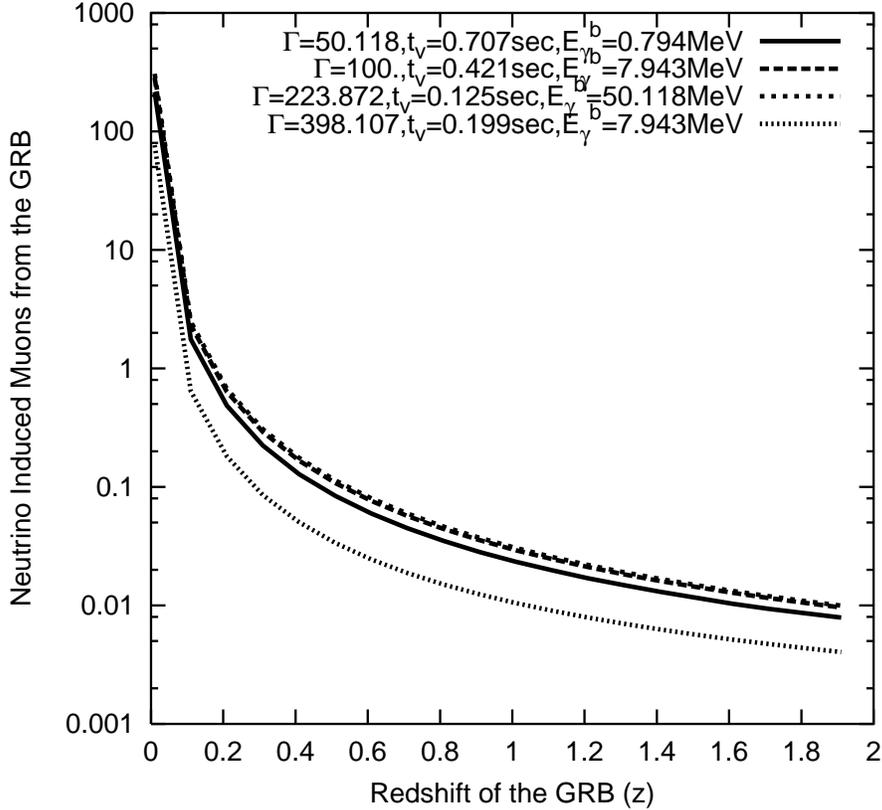,height=11cm} 
\caption{The number of neutrino induced muons in $km^{2}$ area from a GRB have been plotted as a function of redshift $z$ of the GRB. The detector threshold is $E_{th}=100GeV$, the burst is at $\theta=180^{\circ}$ and wind luminosity of the GRB is assumed to be $10^{53}$erg/sec for equipartition parameter $\epsilon_{B}=0.01$. The total energy released in neutrinos is assumed to be $E_{GRB}=10^{53}$erg. Here  $\Gamma$ is the Lorentz factor, $t_v$ is the wind variability time and $E_{{\gamma},MeV}^{b}$ is the observed photon spectral break energy \cite{Guetta}.}
\end{figure}
\begin{figure}
\psfig{figure=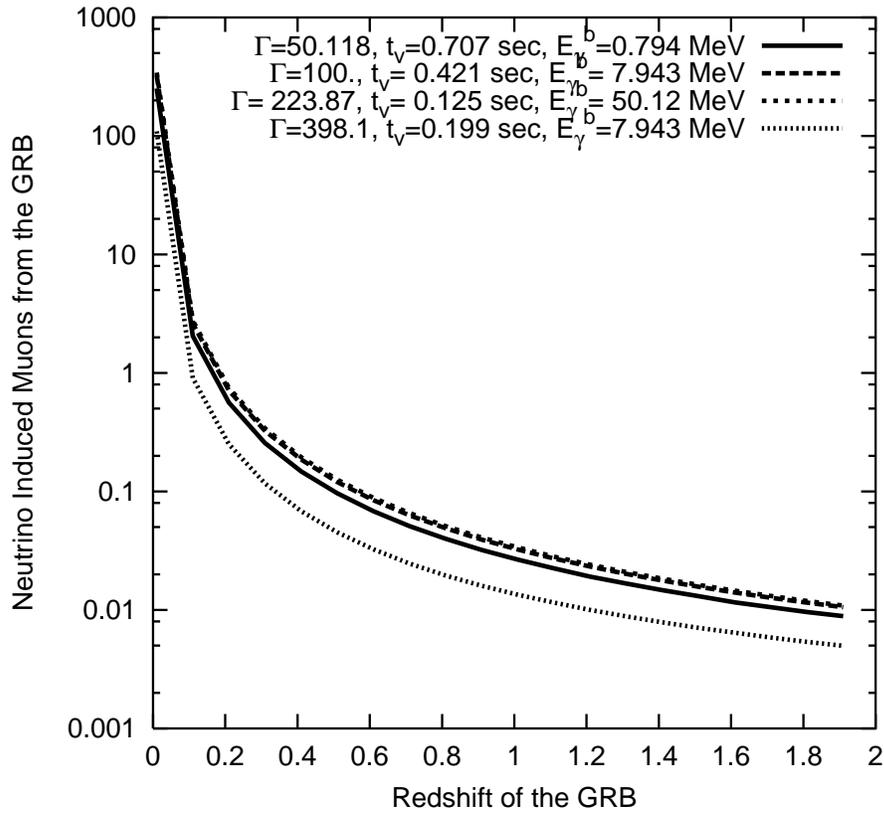,height=11cm}
\caption{All the parameters are same as in Fig.2. only in this case the GRB is
observed at zenith angle $\theta=92.86^{\circ}$ from the South Pole. }
\end{figure}
\begin{center}
\begin{figure}
\psfig{figure=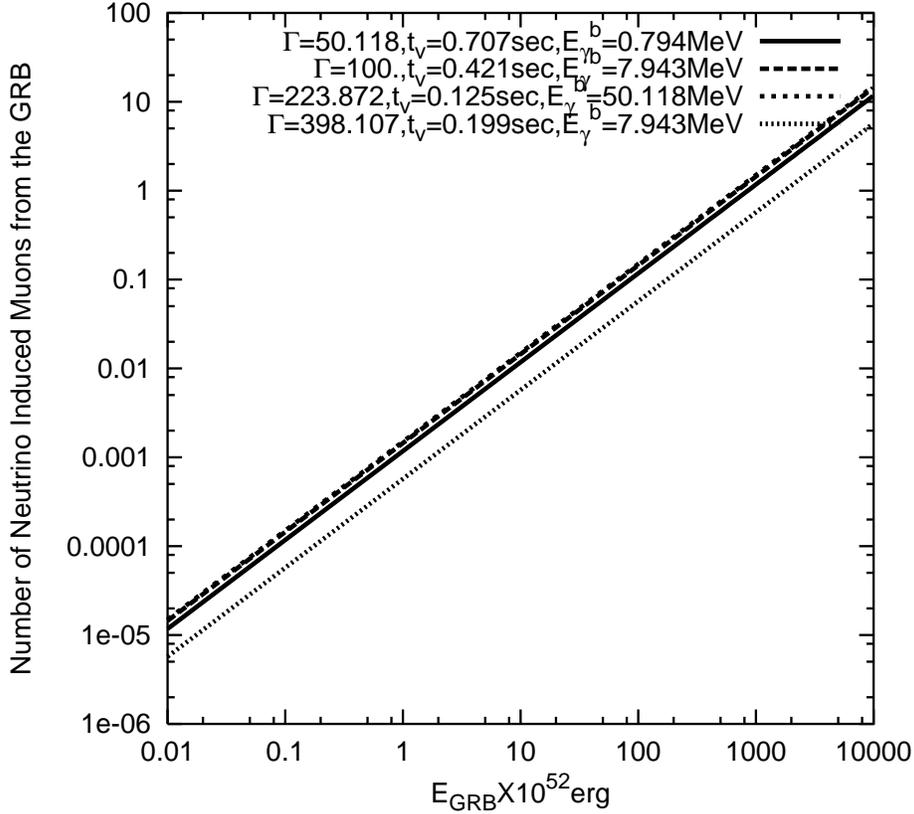,height=11cm} 
\caption{The dependence of number of neutrino induced muons from a GRB in IceCube detector on the total GRB energy $E_{GRB}$ emitted in neutrinos have been shown in this figure. The detector area is $1 km^{2}$ and the GRB is  at
 $\theta=180^{\circ}$. The redshift of the GRB is assumed to be $z=1.5$, wind luminosity $10^{53}$erg/sec and  $\epsilon_{B}=0.01$. The threshold energy of the detector is $E_{th}=100GeV$. The parameters $\Gamma$, $t_{v}$ and $E_{\gamma}^{b}$ are same as in Fig.2. }
\end{figure}
\end{center}

\begin{center}
\begin{figure}
\psfig{figure=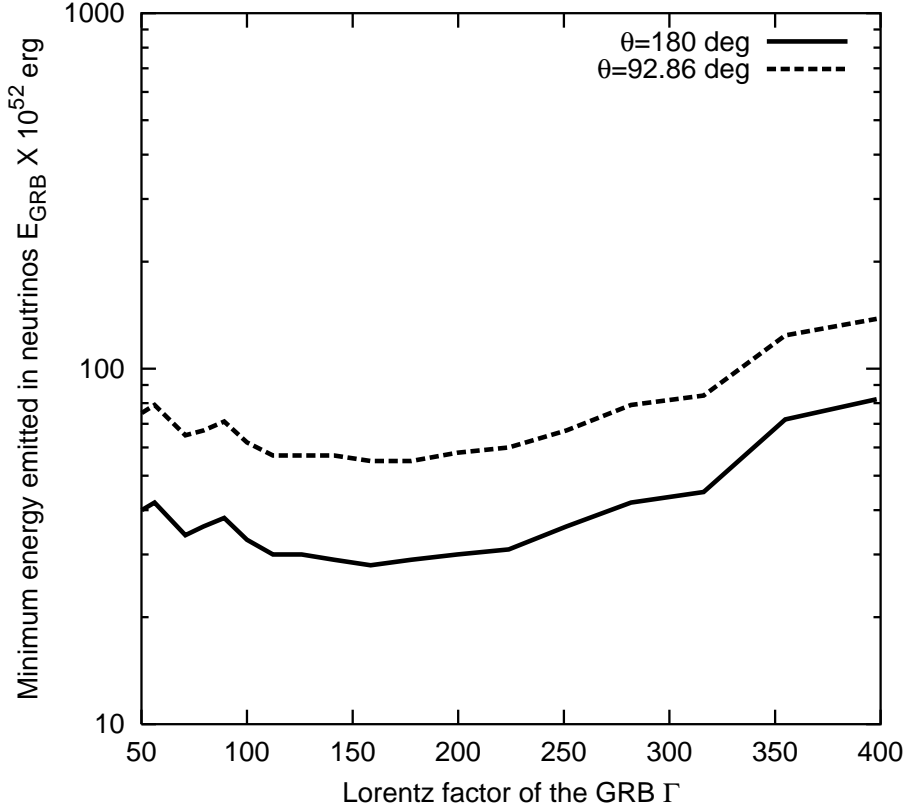,height=11cm}
\caption{In this figure for redshift $z=1.5$ and for zenith angles
$\theta=180^{\circ}$, $\theta=92.86^{\circ}$ the minimum total energy emitted
in neutrinos by the GRB for which the GRB will produce signal above the atmospheric neutrino generated background in $km^{2}$ area detector for minimum Lorentz factor $\Gamma$ has been plotted. The threshold energy of the detector is assumed to be $E_{th}=100 GeV$. The regions above the curves give the allowed regions for positive observations. Here the duration of time ($t_{\nu}$) for neutrino emission from the GRB is assumed to be $1.$second. The wind luminosity $L_{w}$ is assumed to be $10^{53}$ erg/sec and equipartition parameter $\epsilon_{B}=0.01$
. The values of $\Gamma$, $E_{{\gamma}, MeV}^{b}$ and $t_{v}$ sec used to do the calculations are given in TABLE I. }
\end{figure}
\end{center}

\end{document}